\documentclass[10pt]{article}

\usepackage{amsmath}
\usepackage{amssymb}
\usepackage{url}

\usepackage{graphicx}

\usepackage[super]{cite}

\usepackage{color} 

\usepackage{setspace} 
\usepackage[labelfont=bf,labelsep=period,justification=raggedright]{caption}

\makeatletter
\renewcommand{\@biblabel}[1]{\quad#1.}
\makeatother

\date{}

\begin{document}

\begin{flushleft}

  {\Large \textbf{Analysis of the human diseasome reveals phenotype
      modules across common, genetic, and infectious diseases}}
  \\
  Robert Hoehndorf$^{1,\ast}$, Paul N. Schofield$^{2}$, Georgios
  V. Gkoutos$^{3}$
  \\
  \bf{1} Computational Bioscience Research Center and Computer,
  Electrical and Mathematical Sciences \& Engineering 
  Division, King Abdullah University of Science and Technology, 4700
  King Abdullah University of Science and Technology, Thuwal
  23955-6900, Kingdom of Saudi Arabia\\
  \bf{2} Department of Physiology, Development \& Neuroscience,
  University of Cambridge, Downing Street, Cambridge, CB2 3EG, UK
  \\
  \bf{3} Department of Computer Science,
  Aberystwyth University, Llandinam Building, Aberystwyth, SY23 3DB,
  UK
  \\
  $\ast$ E-mail: robert.hoehndorf@kaust.edu.sa
\end{flushleft}

\section*{Abstract}

Phenotypes are the observable characteristics of an organism arising
from its response to the environment. Phenotypes associated with
engineered and natural genetic variation are widely recorded using
phenotype ontologies in model organisms, as are signs and symptoms of
human Mendelian diseases in databases such as OMIM and
Orphanet. Exploiting these resources, several computational methods
have been developed for integration and analysis of phenotype data to
identify the genetic etiology of diseases or suggest plausible
interventions. A similar resource would be highly useful not only for
rare and Mendelian diseases, but also for common, complex and
infectious diseases. We apply a semantic text-mining approach to
identify the phenotypes (signs and symptoms) associated with over
8,000 diseases. We demonstrate that our method generates phenotypes
that correctly identify known disease-associated genes in mice and
humans with high accuracy. Using a phenotypic similarity measure, we
generate a human disease network in which diseases that share signs
and symptoms cluster together, and we use this network to identify
phenotypic disease modules.

\section*{Introduction}

Over the last decade, the rapid emergence of new technologies has
redefined our understanding of the genetic and molecular mechanisms
underlying disease.  For example, we can now identify genetic
predisposition to diseases, and responses to environmental factors,
through a rapidly increasing number of genome-wide association
studies. These studies utilize genetic variation in human populations
to identify sequence variants that predispose some individuals to
common or complex diseases. Such studies also reveal a variety of
differences between disease manifestations.  Application of sequencing
technologies to disease studies has been particularly successful for
genetically-based diseases. For example, full exome sequencing is an
approach that has emerged to identify causative mutations underlying
congenital diseases, and is successfully applied widely \cite{Ng2009,
  Biesecker2010}.  In contrast to genetically based diseases, the
investigation of infectious diseases poses an additional challenge as
it requires not only the understanding of the physiology and
patho-physiology of a single organism, but the investigation of two or
more organisms, their interactions, and the response of one organism
to the other. Similarly, investigations of environmentally-based
diseases require understanding the response of organisms to
environmental influences such as chemicals, radiation or habitat.

For each type of disease (genetically-based, environmental, and
infectious), the genetic architecture of an organism plays a vital
role in the disease manifestation it exhibits, including severity of
symptoms, complications, as well as its response to therapeutic
agents.  A key to gaining an in-depth understanding of the molecular
basis of disease is the understanding of the complex relationship
between the genotype of an organism and the phenotypic manifestations
it exhibits in response to certain influences (genetic, environmental,
or exposure to an infectious agent).  To achieve such a goal, it is
imperative that there is a consistent and thorough account of the
various phenotypes (including signs and symptoms) exhibited by an
organism in response to etiological influences.

To utilize phenotype data for disease studies, information about
Mendelian diseases has been historically well documented in various
formats and, more recently, in electronic resources such as the Online
Mendelian Inheritance in Man (OMIM) \cite{Amberger2011} database and
the Orphanet \cite{Weinreich2008} resource. Both OMIM and Orphanet
provide a catalog of human genes and genetic disorders, and contain a
variety of textual information including patient symptoms and signs.
Ontologies (i.e., structured, controlled vocabularies that formally
describe the kinds of entities within a domain) such as the Human
Phenotype Ontology (HPO) \cite{Robinson2008} have been created in an
attempt to provide a comprehensive controlled vocabulary and knowledge
base describing the manifestations of human diseases, and these
ontologies have been applied to characterize diseases in the OMIM and
Orphanet databases \cite{Koehler2014,
  Hoehndorf2013orphanet}. Additionally, ontology-based analysis of
phenotype data has also been shown to significantly improve the
accuracy of finding disease gene candidates from GWAS data
\cite{Zemojtel2014} and assignation of phenotypes to genes in Copy
Number Variation syndromes \cite{Koehler2014cnv}.

The remarkable conservation of phenotypic manifestations across
vertebrates implies a high degree of functional conservation of the
genes participating in the underlying physiological pathways. Our
increasing ability to identify such functions as well as their role in
human disease using a variety of organisms and approaches, such as
forward and reverse genetics, renders animal models valuable tools for
the investigation of gene function and the study of human disease.
Phenotype information related to model organisms is also being
described using ontologies such as the Mammalian Phenotype Ontology
(MP) \cite{Smith2004}, and data annotated with these ontologies is
being systematically collected and organized in model organism
databases \cite{Bello2012}. The systematic coding of phenotypic and
molecular information related to humans and other model species
facilitates integrative approaches for identifying novel
disease-related molecular information \cite{Driel2006, washington2009,
  Hoehndorf2013orphanet}, prioritizing candidate genes for diseases
based on comparing the similarity between animal model phenotypes and
human disease phenotypes \cite{Hoehndorf2011phenome, mousefinder} as
well as predicting novel drug-target interactions, drug targets and
indications \cite{Hoehndorf2012pharmgkb, Hoehndorf2013drugs, Vogt2014,
  Kuhn2013}.

Extension of these strategies and tools for the study of common and
infectious diseases has been hampered by the lack of an infrastructure
providing phenotypes associated with common and infectious diseases,
and integrating this information with the large volumes of
experimentally verified and manually curated data available from model
organisms.  We have now generated a resource of disease-associated
phenotypes for over 8,000 Mendelian, rare, common and infectious
diseases. The phenotypes and diseases are characterized using
ontologies and interoperate with widely used ontologies used for
describing human and model organism phenotypes \cite{Mungall2010}.
We evaluate our phenotype data against its ability to prioritize genes
for human diseases, and demonstrate that our method yields disease
phenotypes that are comparable to those available from OMIM when
applied to finding candidate genes.  Following validation, we
demonstrate the utility of our resource by revealing closely related
disease modules based on common etiological, anatomical as well as
physiological underpinnings. We make our results freely available at
\url{http://aber-owl.net/aber-owl/diseasephenotypes/} and provide a
visualisation environment for them at
\url{http://aber-owl.net/aber-owl/diseasephenotypes/network/}.




\section*{Results}
We have created a resource of disease-associated phenotypes for
diseases in the Human Disease Ontology (DO). For this purpose, we have
identified co-occurrences between names of diseases (from DO) and
names of phenotypes (from HPO and MP) in abstracts and titles of 5
million articles in Medline.

We employ several different scoring functions to rank the
co-occurrences based on their significance within our corpus. In
particular, we use the normalized pointwise mutual information (NPMI),
T-Score, Z-Score and the Lexicographer's mutual information scores
\cite{Bordag2008} to rank the co-occurrences.  The phenotypes
associated with diseases, scored by our scoring functions, can be
viewed and downloaded at
\url{http://aber-owl.net/aber-owl/diseasephenotypes}.

As our scoring functions associate a value with each identified
co-occurrence between a term referring to a disease class and a term
referring to a phenotype class, we use known gene-disease associations
from the OMIM database to identify a cutoff that maximizes the
potential to prioritize candidate genes of disease based on phenotypic
similarity. For this purpose, we use the PhenomeNET system
\cite{Hoehndorf2011phenome} to systematically compute the semantic
similarity between disease phenotypes and mouse model phenotypes, and
we compare the results against known mouse models of disease from the
Mouse Genome Informatics (MGI) database \cite{Bello2012}, as well as,
using human-mouse orthology, to known gene-disease associations in the
OMIM database. We quantify the predictive power of the phenotype by
computing the area under the ROC curve for predicting gene-disease
associations through phenotype similarity.

\begin{figure}[!htb]
\includegraphics[width=.5\textwidth]{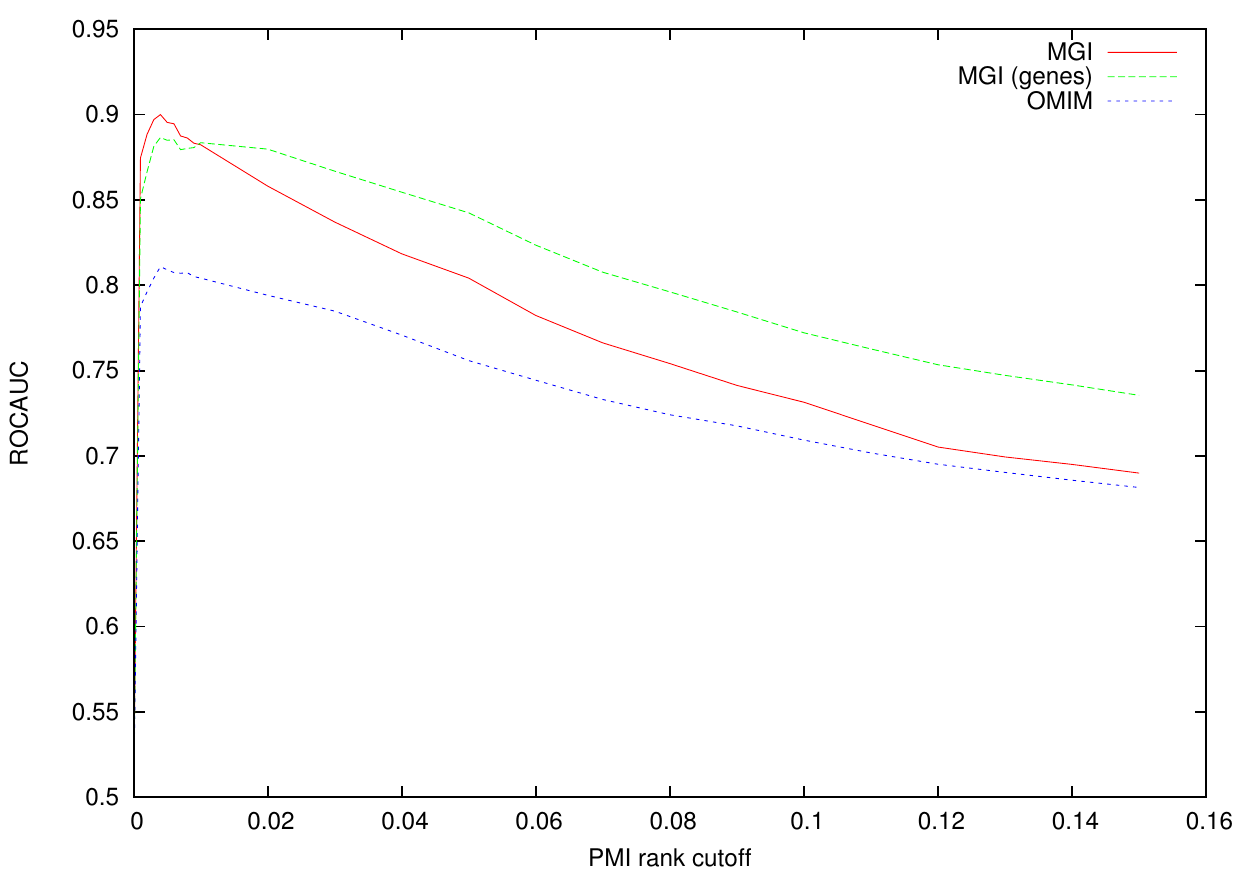}
\caption{\label{fig:pmi-roc}The figure shows the ROCAUC values
  obtained when using different cutoffs for the rank of the pointwise
  mututal information co-occurrence measure. Based on three different
  evaluation datasets, we find that the top 0.4\% ranking
  co-occurrences (NPMI-based) maximize the ROCAUC across our datasets.}
\end{figure}

To standardize the number of phenotypes associated with a disease, we
rank all phenotype-disease associations for each disease by their
normalized pointwise mutual information score. We then perform our
PhenomeNET analysis for an increasing number of phenotypes for each
disease. Figure \ref{fig:pmi-roc} shows the resulting ROCAUC for
varying cutoff values. In particular, we find that using the
top-ranking $0.4\%$ (NPMI-based) of the disease--phenotype associations
maximizes their potential for prioritizing candidate genes using
PhenomeNET, and we use this value as the main cutoff in the remaining
analysis.  Using this cutoff, we have mined phenotypes for 8,672
disease classes in DO, using a total of 12,180 different phenotype
classes from the HPO and MP (7,041 from HPO and 5,139 from MP).

\begin{figure}[!htb]
\includegraphics[width=.5\textwidth]{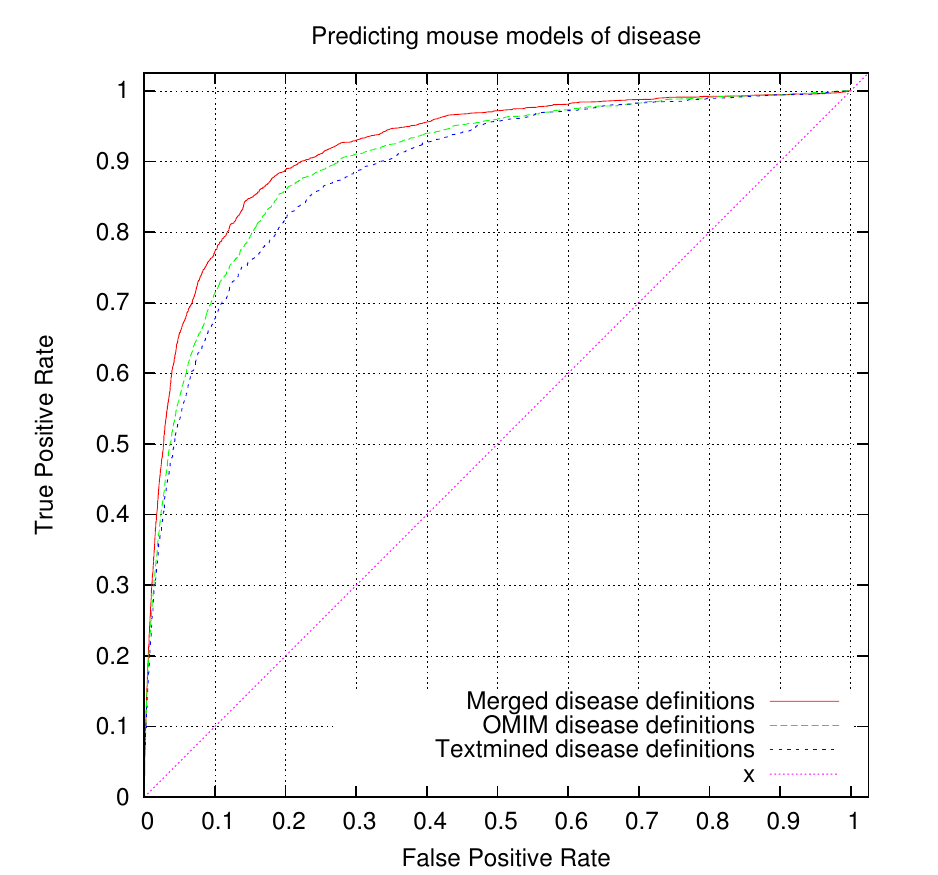}
\caption{\label{fig:roc-mgi-gene}The figure shows the ROC curve for
  cross-species prioritization of disease models using MGI's
  gene-disease association dataset and merging the phenotypes for
  genotypes affecting single genes (ROC AUC: $0.887 \pm .019$
  (text-mined), $0.899 \pm .018$ (OMIM), $0.918 \pm .017$ (merged)).}
\end{figure}

\begin{figure}[!htb]
\includegraphics[width=.5\textwidth]{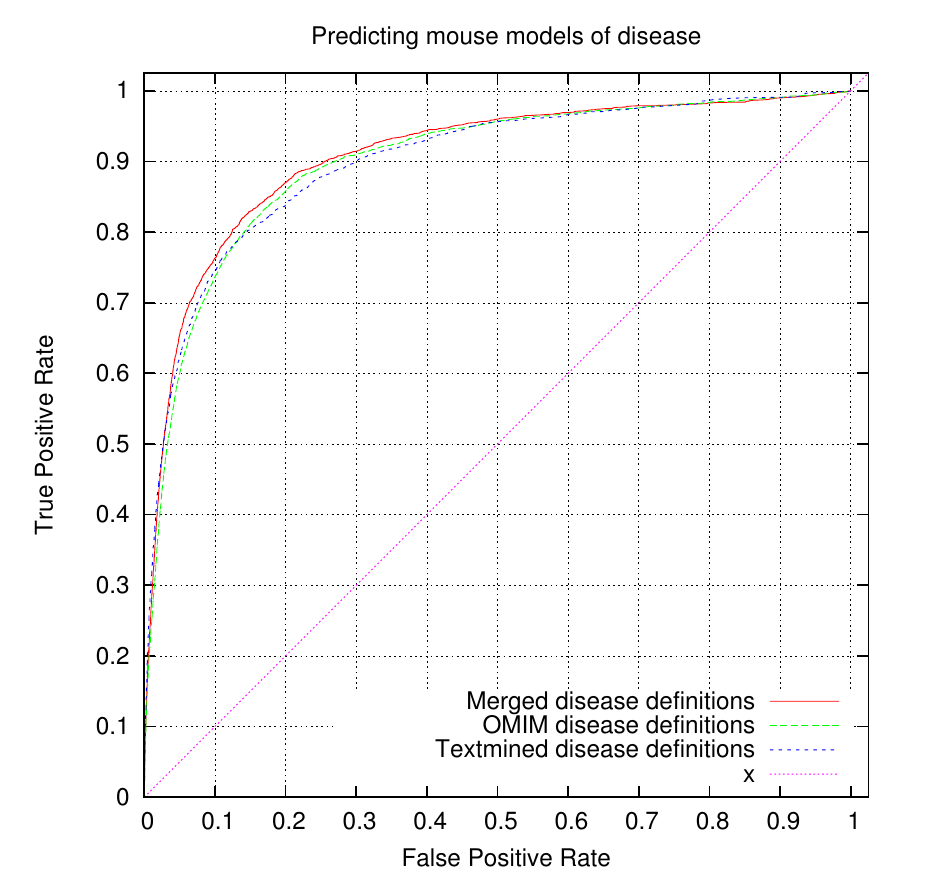}
\caption{\label{fig:roc-mgi}The figure shows the ROC curve for
  cross-species prioritization of disease models using MGI's
  gene-disease association dataset (ROC AUC: $0.900 \pm 0.009$
  (text-mined), $0.900 \pm 0.009$ (OMIM), $0.908 \pm 0.006$
  (merged)).}
\end{figure}

\begin{figure}[!htb]
\includegraphics[width=.5\textwidth]{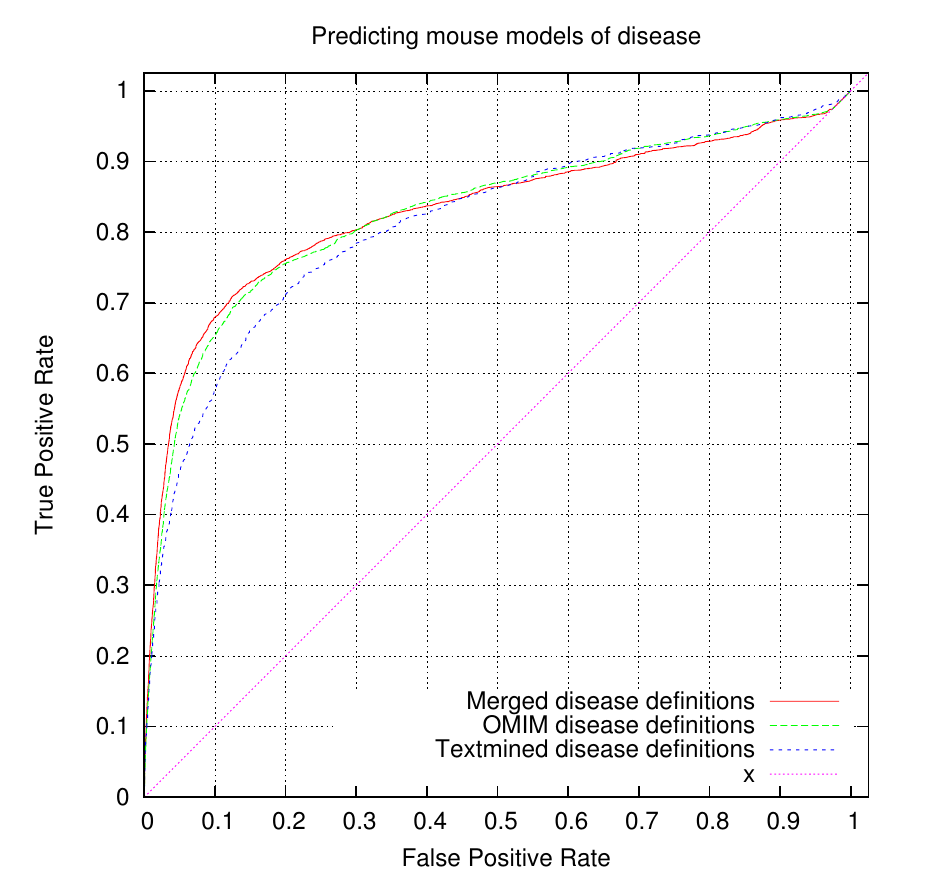}
\caption{\label{fig:roc-omim}The figure shows the ROC curve for
  cross-species prioritization of disease models using OMIM
  MorbidMap's gene-disease association dataset (ROC AUC: $0.811 \pm
  0.011$ (text-mined), $0.828 \pm 0.011$ (OMIM), $0.829 \pm 0.011$
  (merged)).}
\end{figure}

Using only heritable diseases from OMIM, we can demonstrate that our
text-mined phenotypes come close to the phenotypes associated with
OMIM diseases in the HPO database when applied to prioritizing
candidate genes of disease. Figures \ref{fig:roc-mgi-gene},
\ref{fig:roc-mgi} and \ref{fig:roc-omim} show the comparison of the
performance of our text-mined phenotypes with the original OMIM
phenotypes in PhenomeNET. To further test our phenotypes, we have
merged the original OMIM phenotypes with our text-mined phenotypes. In
each case, we could demonstrate an increase in ROCAUC over both our
text-mined phenotypes and the original OMIM phenotypes. In particular,
as can be seen in Figures \ref{fig:roc-mgi-gene},
\ref{fig:roc-mgi} and \ref{fig:roc-omim}, sensitivity of PhenomeNET
gene prioritization increases for the highest ranks when our text
mined phenotypes are merged with the original OMIM phenotypes.

\begin{figure}[!htb]
\includegraphics[width=.5\textwidth]{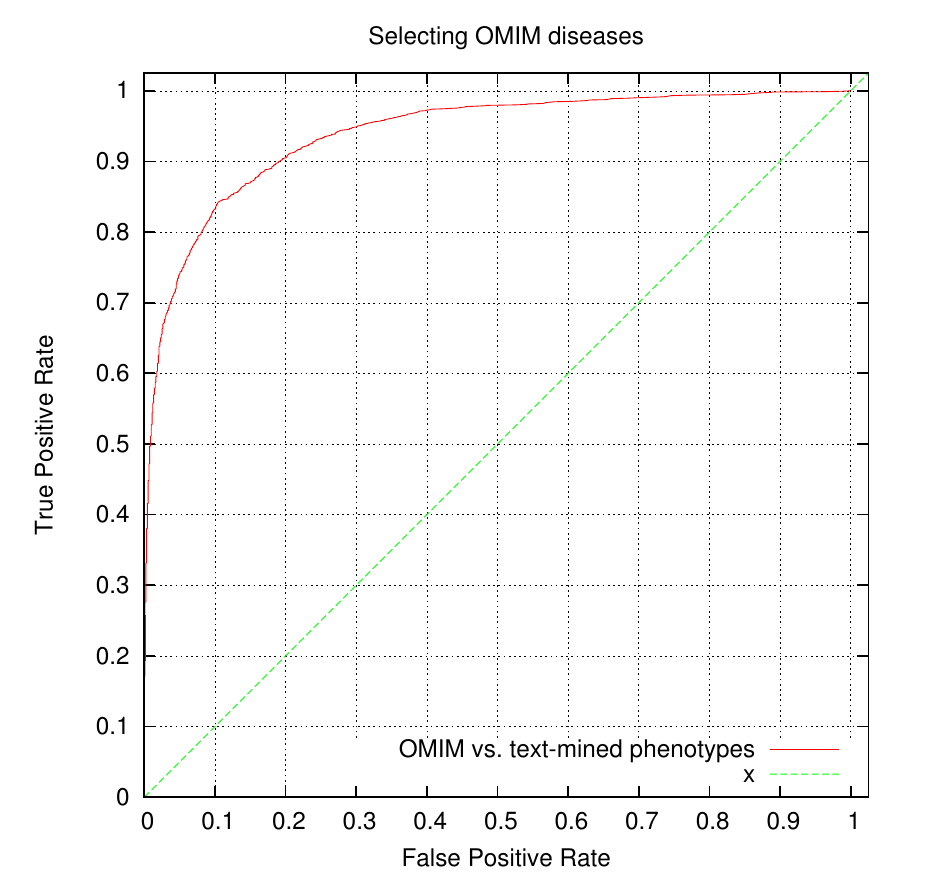}
\caption{\label{fig:roc-omim-recall}The figure shows the ROC curve for
  ranked retrieval of OMIM diseases by semantic similarity to our
  text-mined disease phenotypes (ROC AUC: $0.939 \pm 0.011$).}
\end{figure}

We further evaluate the overlap with OMIM disease definitions, as
characterized by the HPO database. We use two measures to quantify the
overlap. First, we directly compute the set overlap (Jaccard index)
between the HPO phenotypes we have text-mined for each disease and the
HPO phenotypes associated with the disease in the HPO database.  The
average Jaccard index between our disease definitions and the
corresponding OMIM diseases is 0.053 (0.309 when considering the
phenotypes together with all their superclasses).  We also compute the
percentage of coverage of the OMIM phenotypes in our disease
definitions. Using our text-mining approach, we cover on average
17.6\% of the phenotypes in OMIM (46.8\% when considering the
phenotypes together with all their superclasses).
Finally, we compute the semantic similarity between our text-mined
disease definitions and the phenotypes associated with the disease in
HPO, and use ROC analysis to quantify the performance of directly
identifying a matching disease. Figure \ref{fig:roc-omim-recall} shows
the resulting ROC curve.

Using the phenotypes associated with DO diseases, we compute a
pairwise disease--disease similarity based on semantic similarity of
their associated phenotypes. From the resulting similarity matrix, we
generate a disease--disease network based on phenotypes from the
top-ranking 5\% of disease--disease similarity values. The generated
network is shown in Figure \ref{fig:phenome}. For each disease, we
also identify top-level DO categories, and assign node colors in the
network based on the DO categories in which a disease falls.
The disease--disease similarity network can be accessed online at 
\url{http://aber-owl.net/aber-owl/diseasephenotypes/network}.

We also use the disease--disease network to compute phenotypic
homogeneity of diseases within their respective disease category. For
this purpose, for each disease, we sort all other diseases based on
their phenotypic similarity, and identify the ranks at which other
diseases in the same category appear. The results (summarized in Table
\ref{tbl:categories}) are ROCAUC values for each of DO's top-level
categories that quantify how phenotypically similar are the diseases
within the same category.

\section*{Discussion}
\subsection*{Related work}

Associations between phenotypes, signs and symptoms on one side and
diseases on the other have been used to gain insights into the modular
nature and network structure of human diseases and drug indications
\cite{Driel2006, Zhou2014, Xu2013}. In prior work, text-mining has
been used based on labels of diseases and labels of phenotypes (signs
and symptoms) \cite{Xu2013}, or the identifiers of the Medical Subject
Headings (MeSH) Thesaurus \cite{Nelson2004} that are associated with
article citations in Pubmed, have been used to identify associations
between disease and phenotype.  In general, the resulting
disease--phenotype associations have been evaluated based on their
ability to reveal or explain perceived clusters of diseases
\cite{Driel2006}, group diseases with known common etiology together
\cite{Zhou2014, Xu2013}, based on gold standard comparison and
clustering for common drug targets \cite{Xu2013}.

One fundamental question that has not been answered by any of the
prior approaches has been what kind of evidence or support would be
required to consider a disease--phenotype association as
``correct''. This is a fundamental challenge in any kind of phenotype-
or symptom-based characterization of disease. Most diseases have
cardinal signs and symptoms which will always be associated with a
disease. However, a large number of signs and symptoms for a disease are not
always present but rather occur with varying frequency, and even very
rare manifestations may prove to be highly useful in the context of
differential diagnosis. In our evaluation, we provide a quantifiable
measure through comparison against experimental data which can be used
to determine -- and maximize -- the {\em utility} of our text-mined
disease--phenotype associations. We therefore provide an objective
measure that can be used to determine how applicable a set of
disease--phenotype associations are to a particular scientific
question -- in our case, identifying candidate genes for
diseases of genetic origin.

One main limitation of our evaluation is that it is limited to
genetically-based diseases, while the majority of diseases in the DO
is not genetically-based. Other approaches, such as clustering
diseases based on similarity and identifying meaningful, well-known
clusters \cite{Driel2006, Zhou2014}, or comparison with known drug
indications \cite{Xu2013}, can evaluate the biological validity of
generated associations, but often cannot quantify the results.

\subsection*{Novel candidate genes based on text-mined phenotypes}
Through our approach, we do not only obtain phenotypic
characterization of common and infectious diseases, but we have also
obtained novel phenotype associations for genetically based diseases in
OMIM for which currently no phenotypic characterization exists either
in the HPO annotations or as a clinical synopsis in OMIM.

The HPO database contains phenotype annotations for 9,286 OMIM entries
(genes and diseases). Through the DO--OMIM mappings and our method, we
obtain phenotypes for 1,683 OMIM entries, 115 of which have no
phenotype annotations in HPO or an associated clinical synopsis in
OMIM. For example, {\em Halo Nevi} ({\em Leukoderma acquisitum
  centrifugum of sutton}, {\tt OMIM:234300}), a dermatological
condition in which melanocytes are destroyed by CD8+ cytotoxic T
lymphocytes \cite{Mundinger2014}, has currently no clinical synposis
in OMIM and consequently no associated phenotypes in the HPO database,
while we identify several phenotypes, including {\em Irregular
  hyperpigmentation} ({\tt HP:0007400}), {\em abnormal dermal
  melanocyte morphology} ({\tt MP:0009386}) and {\em Progressive
  vitiligo} ({\tt HP:0005602}) as phenotypes, all of which are known
to be associated with halo nevi \cite{Kopf1965}.

\begin{figure}[!htb]
\includegraphics[width=.5\textwidth]{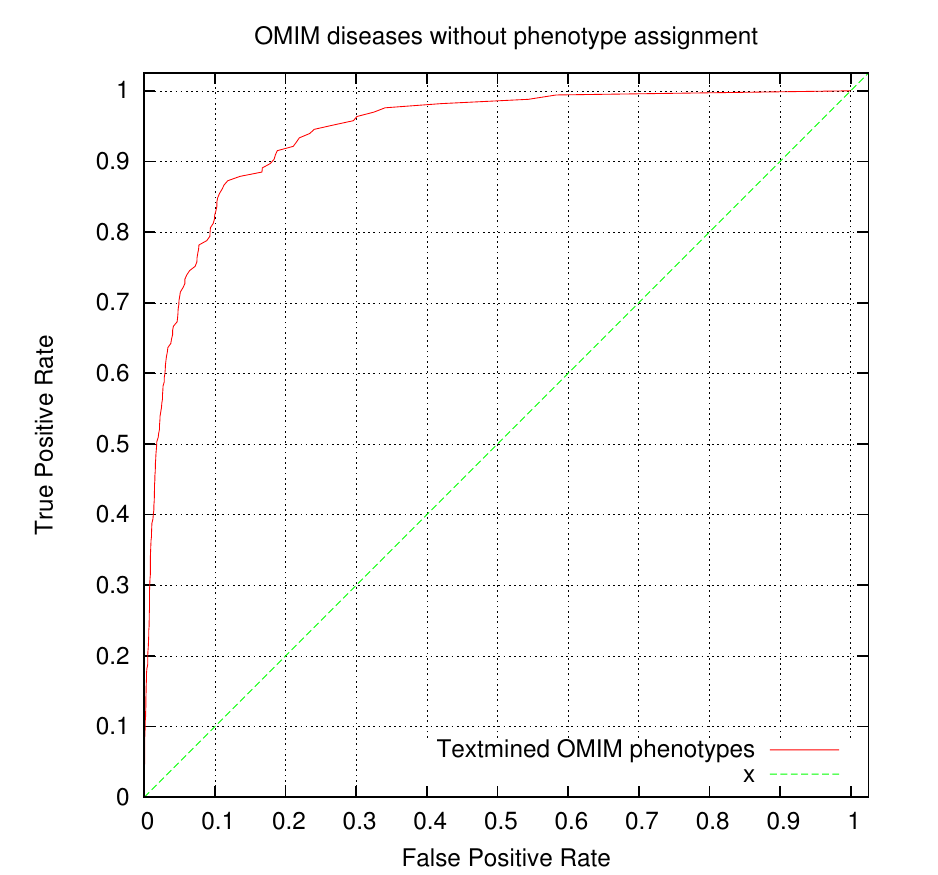}
\caption{\label{fig:roc-no-def-in-omim}The figure shows the ROC curve for
  ranked retrieval of MGI disease models by semantic similarity to 
  text-mined phenotypes of diseases without clinical synopsis in OMIM
  (ROC AUC: $0.940 \pm 0.018$).}
\end{figure}

For these 115 diseases, 167 disease models are known in the mouse. We
can prioritize the correct model with ROCAUC of $0.940 \pm 0.018$ for
this set of 115 diseases (Figure \ref{fig:roc-no-def-in-omim}).

\subsection*{Exploring disease--disease similarities: revealing the
  modular nature of disease}

\begin{figure}[!htb]
\centerline{\includegraphics[width=.8\textwidth]{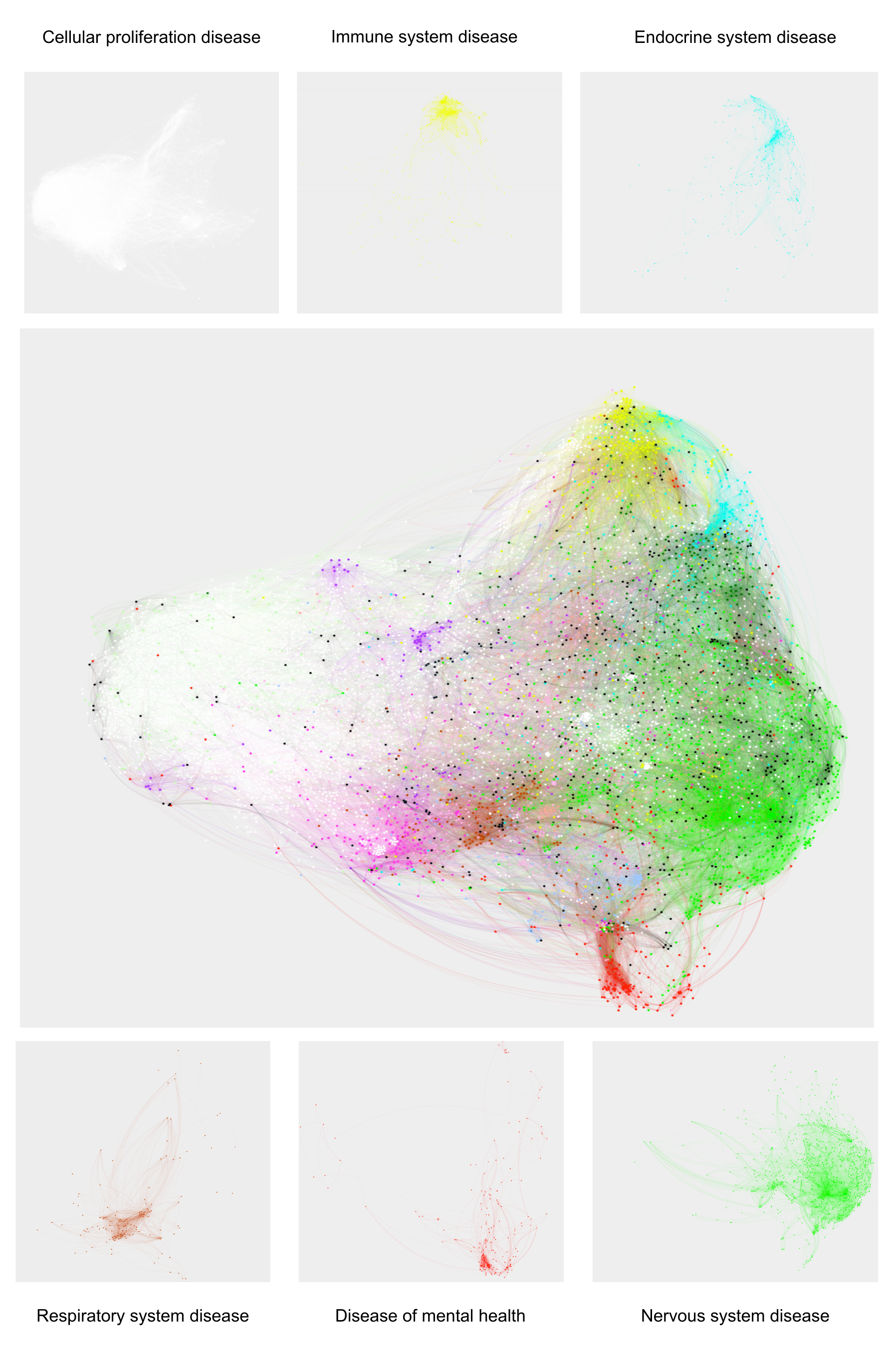}}
\caption{\label{fig:phenome}An overview over the disease--disease
  similarity network generated by our approach as well as six disease
  modules obtained by filtering for disease categories in DO. The
  graph is based on a force-directed layout using the similarity
  between diseases as attraction force.}
\end{figure}

In Figure \ref{fig:phenome}, we show the relationships between
common, genetic, infectious and environmental diseases. Each node in
the network represents a disease and is coloured according to its
corresponding top-level disease class in DO. Using this similarity
network, it is clear that diseases of different systems and
pathological processes can be separated on the basis of phenotypic
relatedness.  DO classifies both by anatomical site or system, and by
general pathology, and for each of the classifications, despite these
different criteria, we find that diseases within one category cluster
together on the basis of phenotypic relatedness alone. In Figure
\ref{fig:phenome}, we highlight different upper-level disease
categories from DO, including neoplasias, immune diseases, respiratory
diseases, mental health diseases, endocrine, and nervous system
disease.  While many diseases cluster tightly within their group, as
expected, several diseases show significant phenotypic relations to
different areas or systems, and we see many examples in which a
disease clusters predominantly within one part of the DO-defined area
but has more distant relationships with others.  For example, {\em
  phaeochromocytoma} is associated with other adrenal tumors in the
class of ``neoplasia'' but also with {\em adrenal gland
  hyperfunction}, {\em adrenal cortex disease} from the category of
endocrine diseases, and {\em hypertension} in the cardiovascular
disease category. {\em Hyperprolactinaemia} has relations to a cluster
of pituitary tumours and, as expected, to {\em prolactinomas}
(neoplasms), {\em acromegaly} (physical disorders) and {\em
  hypogonadotropism} (reproductive system), and, more distantly, {\em
  psychological dyspareunia} (mental).

\begin{figure}[!htb]
\includegraphics[width=.5\textwidth]{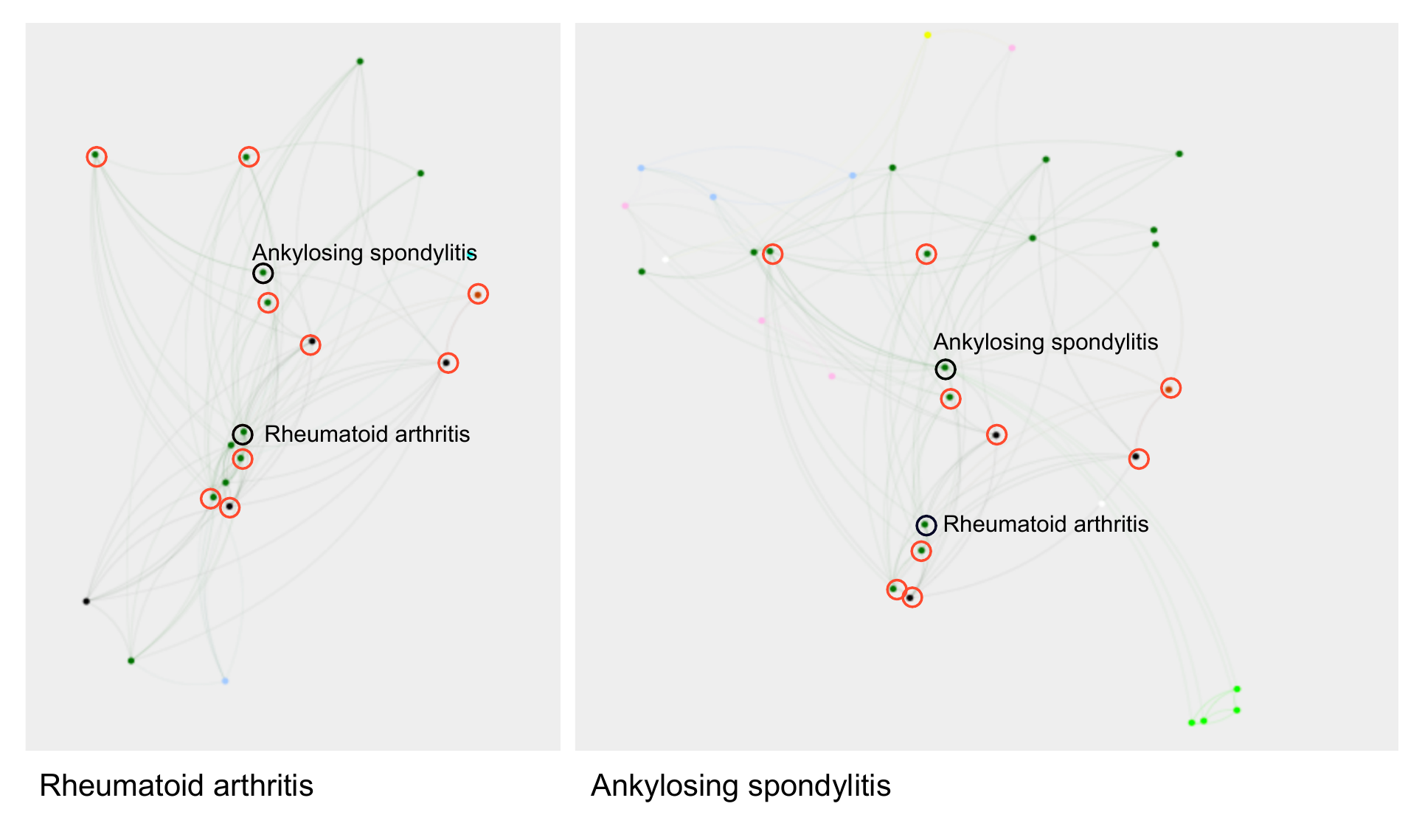}
\caption{\label{fig:ankylosing}The networks surrounding ankylosing
  spondylitis and rheumatoid arthritis. Diseases common to both
  networks are ringed in red (see main text).}
\end{figure}

We can also identify phenotypically-defined ``footprints'' for disease
groups which show overlapping phenotypic similarity. For example
comparing the disease networks centered on rheumatoid arthritis (RA)
and ankylosing spondylitis (AS), it is clear that the two are quite
closely related to the same group of inflammatory diseases (Figure
\ref{fig:ankylosing}). However, a close phenotypic relationship to
rheumatic fever and rheumatoid lung disease is missing from the
ankylosing spondylitis-centered network, and uveitis is missing from
that of rheumatoid arthritis. Uveitis forms one of the diagnostic
features of ankylosing spondylitis, and some of the most common
diseases that result in uveitis are ankylosing spondylitis and
juvenile rheumatoid arthritis \cite{Amor1990}. Acute anterior uveitis
is the most common extra-articular feature of AS, occurring in
25\%--40\% of patients at some time in the course of their disease. AS
and uveitis share an association with HLA-B haplotypes, indicating the
possible existence of a modular phenotype linking these inflammatory
diseases with a common genetic etiology
\cite{Sampaio2006}. Interestingly Felty's syndrome with
thromobocytopenia and vasculitis, present in the network, is also
associated both with the spondylopathies and the HLA-B haplotypes, and
there are suggestions of a relationship between ankylosing spondylitis
and vascular inflammatory disease in addition to the common cardiac
effects (Figure \ref{fig:ankylosing}) \cite{Palazzi2010}.

\begin{figure}[!htb]
\centerline{\includegraphics[width=.8\textwidth]{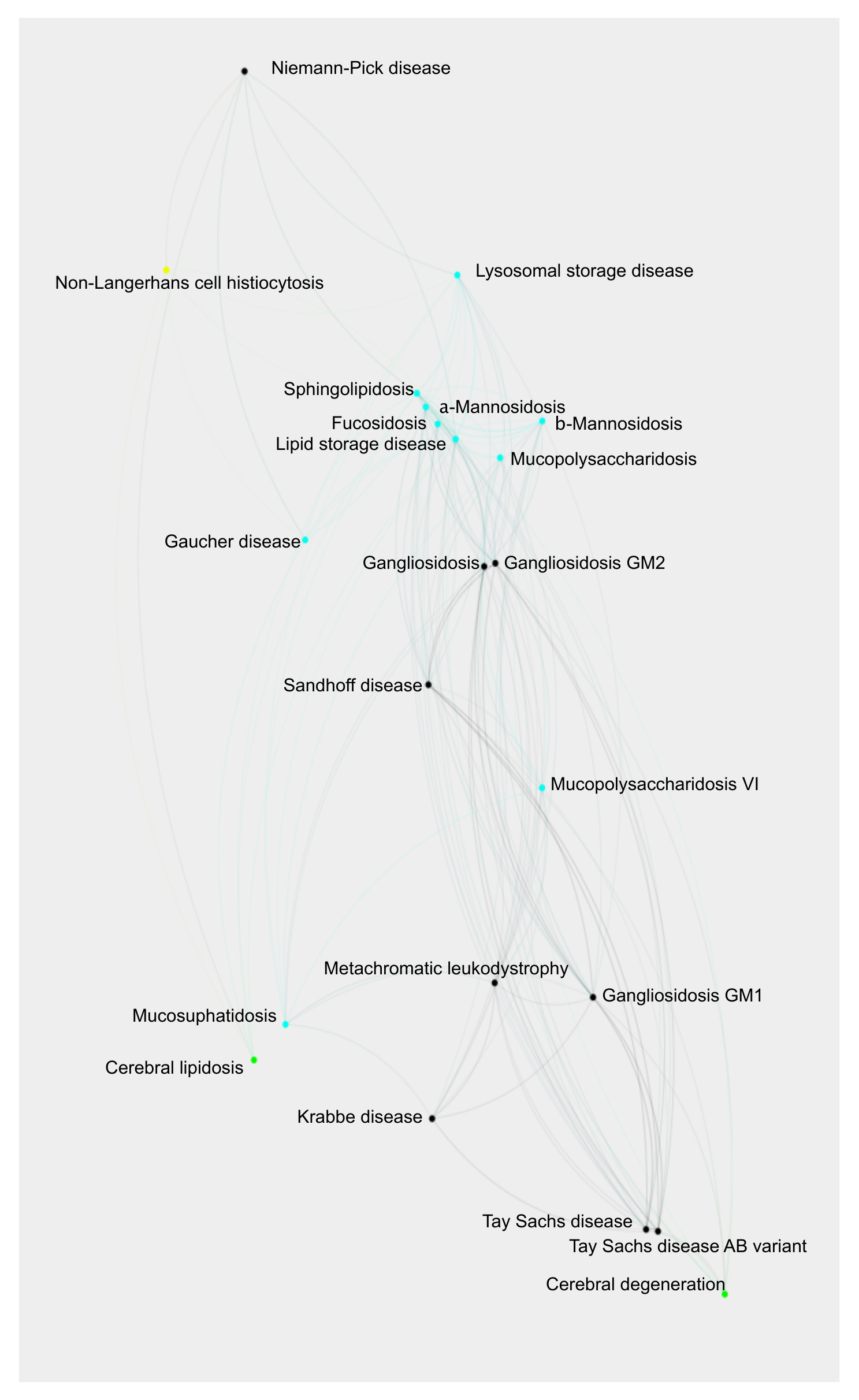}}
\caption{\label{fig:storage}The sub-network around lysosomal storage diseases.}
\end{figure}

Another example of phenotype-defined disease groups are the lysosomal
storage diseases. All cells contain lysosomes which contain soluble
acid hydrolases whose role is to process a wide range of
substrates. Failure to perform this function results in accumulation
of lysosomal accumulation of unmetabolized proteins lipids and
carbohydrates, which are the primary cause of disease through their
effects on cellular metabolism. The pathways by which these
accumulations exert their pathological effects are only just becoming
understood, but they display an extensive range of disease symptoms
with central neurological involvement and a wide range of peripheral
phenotypes with very variable individual manifestation
\cite{Vitner2010}.  Figure \ref{fig:storage} shows the relationships
between sphingolipidoses, mucopolysaccharidoses, and
oligosaccharidoses, demonstrating a coherent phenotypic disease
footprint for this wide range of lysosomal storage disorders. This
striking clustering is similar in type to that seen in the
ciliopathies \cite{Hildebrandt2011} where a range of related
phenotypes reflect lesions in a collection of molecules involved in
different aspects of cilium assembly or function, which, along with
other examples, lead Oti and Brunner \cite{Oti2007} to postulate the
existance of common functional modules underlying the phenotypic
profiles of diseases.  Phenotypic annotation for these diseases
benefits not only from organismal level description but in many cases
molecular annotations such as {\em Abnormality of proteoglycan
  metabolism} ({\tt HP:0004355}). While this alone does not account
for the clustering, the increased depth available for these diseases
greatly improves the quality of the network
associations. 

\begin{figure}[!htb]
\centerline{\includegraphics[width=.5\textwidth]{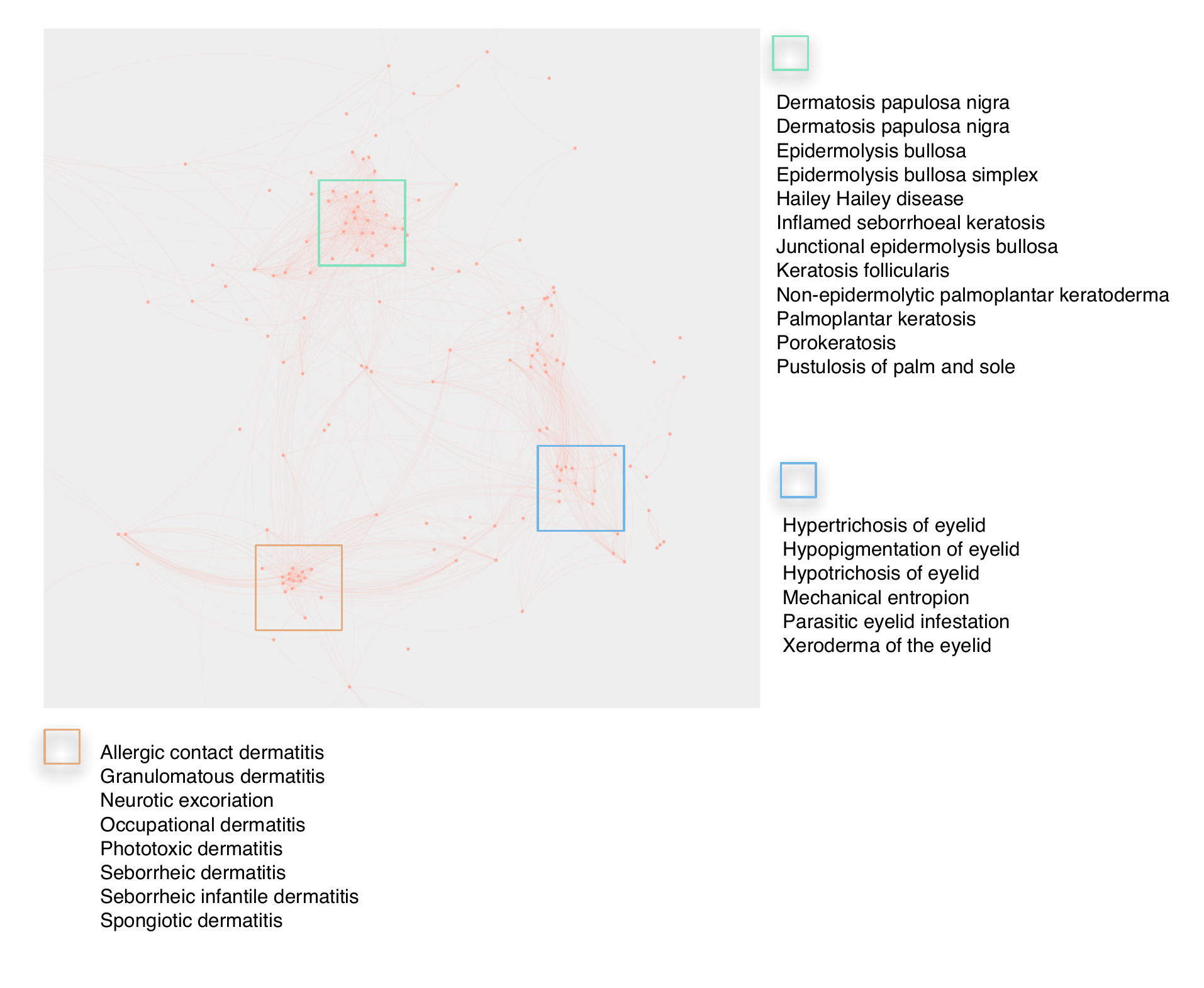}}
\caption{\label{fig:dermo}Network of dermatological disorders.  We
  highlight three phenotypic sub-modules. The first contains the
  inflammatory skin diseases, the second diseases clustered by anatomy
  (the eyelid), and the third a combination of benign proliferative
  disorders and those of keratinisation, all of which here show
  overlapping phenotypes.}
\end{figure}

Within disease modules, we find the separation of diseases by both
anatomical site and pathology. For example, in Figure \ref{fig:dermo},
the integumentary diseases form a distinct group and show clear
clustering of inflammatory skin diseases, such as {\em seborrheic
  dermatitis} and {\em granulomatous dermatitis} along with {\em
  neurotic excoriation} which itself often involves inflammation as a
consequence of compulsive ``skin picking''. An additional cluster is
evident which includes benign proliferative disorders and those of
keratinisation, together with bullous diseases, such as {\em
  epidermolysis bullosa}, themselves involving acantholysis. Finally,
there is a group of diseases of the eyelid, ranging from mechanical
lesions to parasitic disease. {\em Alopecia telogen effluvium}, {\em
  alopecia areata}, {\em alopecia universalis} and {\em follicular
  mucinosis} similarly cluster together, all diseases involving hair
follicles and causing hair loss \cite{Bolognia2012}.

Unsurprisingly, we also find (see Table \ref{tbl:categories}) that
diseases classified by anatomical site or system (e.g., thoracic
diseases, respiratory diseases) exhibit higher phenotypic homogeneity
than diseases classified by their pathological mechanism (e.g.,
infectious diseases, genetic diseases). In particular, we observe that
narrowly defined disease categories such as {\em thoracic disease} or
{\em respiratory disease} exhibit a high phenotypic homogeneity; broad
categories such as all the infectious diseases, on the other hand, are
relatively heterogeneous. However, all of DO's top-level categories
cluster significantly based on their phenotypic similarity, and
diseases falling into more specific DO categories (such as lysosomal
storage diseases) cluster closely as well, demonstrating that not only
Mendelian diseases form disease modules \cite{Driel2006, Oti2007}, but
also common diseases.

\subsection*{Conclusions}
Exploring diseases through their associated phenotypes associated has
major applications for biomedical research, and several studies have
primarily relied on disease phenotypes to reveal functional disease
modules \cite{Oti2007, Driel2006, Zhou2014}, candidate genes of
disease \cite{Chen2012, Hoehndorf2011phenome}, prioritize genes in
GWAS studies \cite{Robinson2013}, and investigate drug targets and
indications \cite{Vogt2014, Hoehndorf2013drugs, Campillos2008,
  Gottlieb2011}.  While the majority of these investigations have been
focused on genetic diseases, application of similar methods may lead
to novel insights into the patho-biology of common and infectious
diseases as well.

\section*{Materials and Methods}
\subsection*{Ontologies and vocabularies}

We use the Human Phenotype Ontology (HPO) \cite{Robinson2008} and the
Mammalian Phenotype Ontology (MP) \cite{Smith2004} as vocabularies
that provide terms referring to phenotypes, signs and symptoms
associated with diseases. Additionally, the MP is used to describe
mouse model phenotypes \cite{mgi}, and we rely on comparison to mouse
model phenotypes for the evaluation of our approach.

We use the Human Disease Ontology (DO) \cite{do} as an ontology of
diseases. The DO contains a rich classification of rare and common
diseases, and spans heritable, developmental, infectious and
environmental diseases.
All ontologies were downloaded from the OBO Foundry website
\cite{Smith2007} on 2 July 2013.

\subsection*{Semantic mining with Aber-OWL: Pubmed}
We make use of the Aber-OWL: Pubmed infrastructure to semantically
mine Medline abstracts. Aber-OWL: Pubmed
(\url{http://aber-owl.net/aber-owl/pubmed/}) consists of an ontology
repository, a reasoning infrastructure capable of performing OWL-EL
reasoning over the ontologies in the repository, a fulltext index of
all Medline 2014 titles and abstracts as well as all Pubmed Central
articles, and a search interface.  Aber-OWL: Pubmed uses an Apache
Lucene (\url{http://lucene.apache.org}) index to store the
articles. Before indexing, every text is processed using Apache
Lucene's English language standard analyzer which tokenizes the text,
normalizes text to lower case, and applies a list of stop words.

To identify documents which contain references to a disease or
phenotype term, we first limit our search to Medline abstracts and
treat documents as consisting of a title and the abstract. We then
limit our corpus to documents in which at least one term from a
phenotype ontology (HPO or MP) or the DO occurs. As a result of this
filtering step, we use a corpus consisting of 5,164,316 documents.

We use the information in ontologies together with the Aber-OWL
reasoning infrastructure to identify the set of terms referring to a
disease or phenotype. For this purpose, we first identify all labels
and synonyms $Lab(C)$ associated with a class $C$ in an ontology. We
then define the set of terms $Terms(C)$ referring to a class $C$ as:
\[
Terms(C) := \{ x | x \in Lab(S) \land S \sqsubseteq C\}
\]
According to this definitions, the set $Terms(C)$ refers to the set of
labels and synonyms of $C$ or any subclass of $C$, as inferred using
the automated reasoner employed by the Aber-OWL infrastructure.

To identify the number of documents in which a disease or phenotype
term occurs, we construct a Lucene query based on $Terms(D)$ and
$Terms(P)$ in which we concatenate each member of $Terms(D)$ or
$Terms(P)$ using the {\tt OR} operator: $\bigvee_{x \in Terms(D)}x$
and $\bigvee_{x \in Terms(P)}x$.  As a result, the Lucene query will
match any document (title or abstract) that contains a label or
synonym of a class $D$ or $P$. To identify the number of documents in
which $D$ and $P$ occur together, we concatenate both queries using
the {\tt AND} operator: $\bigvee_{x \in Terms(D)}x \land \bigvee_{x
  \in Terms(P)}x$.

We use $Docs(q)$ to refer to the set of documents satisfying the query
$q$, $n_D$ to refer to the number of documents in which a term
referring to disease $D$ occurs, $n_P$ to refer to the number of
documents in which a term referring to a phenotype $P$ occurs, and
$n_{DP}$ to refer to the number of documents in which both a term
referring to $D$ and a term referring to $P$ occurs:
\begin{equation}
n_D = |Docs(\bigvee_{x \in Terms(D)}x))|
\end{equation}
\begin{equation}
n_P = |Docs(\bigvee_{x \in Terms(P)}x))|
\end{equation}
\begin{equation}
n_{DP} = |Docs(\bigvee_{x \in Terms(D)}x \land \bigvee_{x \in Terms(P)}x)|
\end{equation}
$n_{tot}$ is the total number of documents in our corpus (5,164,316).

We compute several co-occurrence measures \cite{Bordag2008} to
determine whether a co-occurrence between a term referring to a
phenotype and a term referring to a disease is significant. In
particular, we compute the Normalized Pointwise Mutual Information
(NPMI), T-Score, Z-Score, and Lexicographer's Mutual Information (LMI)
measures \cite{Bordag2008}:
\begin{equation}
NPMI(D,P) = \frac{\log(\frac{n_{DP}\cdot n_{tot}}{n_D \cdot n_P})}{-\log(\frac{n_{DP}}{n_{tot}})}
\end{equation}
\begin{equation}
TScore(D,P) = \frac{n_{DP} - \frac{n_D \cdot n_P}{n_{DP}^2}}{\sqrt{n_{DP}}}
\end{equation}
\begin{equation}
ZScore(D,P) = \frac{n_{DP} - \frac{n_D \cdot
    n_P}{n_{DP}^2}}{\sqrt{\frac{n_{D}\cdot n_{P}}{n_{tot}^2}}}
\end{equation}
\begin{equation}
LMI(D,P) = n_{DP} \cdot \log \frac{n_{tot}\cdot n_{DP}}{n_D \cdot n_P}
\end{equation}

We use NPMI as our primary scoring function for phenotype--disease
associations; the other scoring functions are pre-computed and made
available for further analysis on our website. Based on a score for a
co-occurrence, we can sort phenotype associations for each disease
based on decreasing score values. Using this sorted list, we then
compute a rank for an association such that the highest-scoring
association for a disease is on rank $0$. We use this ranking based on
the NPMI score to determine a rank-based cut-off; in particular, we
set a cut-off based on highest-scoring $p$ percent of the
associations. Using the rank as cut-off instead of raw score value
allows comparison across multiple diseases, as each disease will have a
the same number of phenotypes associated, independent of the actual
value of the score.

\subsection*{Semantic similarity}
We use the PhenomeNET system to compute the semantic similarity
between disease phenotypes and mouse model phenotypes. PhenomeNET
\cite{Hoehndorf2011phenome} integrates multiple species-specific
phenotype ontologies into a single structure in which classes are
related based on their formal definitions \cite{Mungall2010}. For
example, the HPO class {\em Tetralogy of Fallot} ({\tt HP:0001636})
will become a subclass of the MP classes {\em ventricular septal
  defect} ({\tt MP:0010402}), {\em overriding aortic valve} ({\tt
  MP:0000273}) and {\em abnormal blood vessel morphology} ({\tt
  MP:0000252}), among others, based on the definitions that were
developed for the classes in both ontologies.  As a consequence of
this cross-species integration, it becomes possible to directly
compare phenotypes and sets of phenotypes across species using
approaches based on semantic similarity \cite{Couto2009}.

To compare sets of phenotypes (either associated with a disease, or
observed in a mouse model), we use the set-based simGIC measure
\cite{Pesquita2008}. The simGIC measure is based on the Jaccard index
weighted by information content of a class within the corpus
consisting of mouse models and diseases:
\begin{equation}
  simGIC(P,R) = \frac{\displaystyle\sum\limits_{x\in Cl(R) \cap
      Cl(P)}IC(x)}{\displaystyle\sum\limits_{y\in Cl(R) \cup
      Cl(P)}IC(y)}
\end{equation}
where $Cl(X)$ is the smallest set containing $X$ and which is closed
against the superclasses relation (i.e., $Cl(X) = \{ a | a \in X \lor
\exists y(y \in X \land a \sqsupseteq y)\}$). $IC(x)$ is the
information content of a class $x$ within the corpus of mouse models
and diseases (i.e., $IC(x) = - log (p(x))$).

\subsection*{Evaluation}
We evaluate our text-mined disease phenotypes by comparing the semantic
similarity between the disease phenotypes and mouse model
phenotypes. We assume that semantic similarity over phenotype
ontologies (``phenotypic similarity'') is indicative of a causal
relation between the mutation underlying the mouse phenotypes and the
disease. For this purpose, we compare the results against three
curated datasets of known gene-disease associations for heritable
diseases. All other associations are treated as negative instances for
the purpose of the evaluation.  As evaluation datasets, we use two
sets of gene-disease associations: the gene-disease associations from
OMIM's MorbidMap \cite{Amberger2011} and the genotype-disease
associations from the MGI database \cite{mgi}. We generate the third
evaluation dataset by taking the genotype-disease associations from
the MGI database, filtering by single gene knockouts, and merging all
phenotypes associated with one gene. As a result, our third dataset
consists of gene-disease associations. We refer to the three
evaluation datasets as ``OMIM'', ``MGI'' and ``MGI (genes)'',
respectively.

We use receiver operating curve (ROC) analysis to evaluate and
quantify the predictive power of the text-mined disease phenotypes. A
ROC curve is a plot of the true positive rate of a classifier as a
function of the false positive rate, and the area under the ROC curve
(ROCAUC) is a quantitative measure of a classifier's quality
\cite{Fawcett2006}.  We report the ROC AUC values together with an
estimate of the $95\%$ confidence interval \cite{Birnbaum1957}: we use
$\sigma_{max}^2 = \frac{AUC(1-AUC)}{min\{m,n\}}$, with $m$ and $n$
being the number of positive and negative instances in the evaluation
dataset, and then use $AUC \pm 2\sigma$ as an estimate of the 95\%
confidence interval. 

\subsection*{Interface and visualization}
The web interface was written in Groovy (backend) and Javascript
(frontend). The disease network was visualized using the Gephi graph
visualization tool \cite{gephi2}, and the disease network browser was
generated with Gephi's Sigma\.js export plugin
(\url{http://blogs.oii.ox.ac.uk/vis/}). All graphs are visualized
using a force-directed layout.

\section*{Acknowledgments}
We thank Prof. John Sundberg for his helpful guidance on skin
disorders, and Prof. Peter Vogel for his insightful comments on the
lysosomal storage disorders.

\section*{Author Contributions}
RH, PNS and GVG conceived of the study, interpreted the results, wrote
and reviewed the final version of the manuscript. RH performed the
analysis.

\section*{Additional Information}
No special funding was received for this study, and we declare that we
have no competing financial interests in relation to this work.

\clearpage

\bibliographystyle{naturemag}

\clearpage
\section*{Figures}

\clearpage
\section*{Tables}

\begin{table}[!htb]
\centering
\begin{tabular}{|l|r|}
\hline{\bf Disease category} & {\bf ROCAUC} \\
\hline&\\
bacterial infectious disease ({\tt DOID:104}) & $0.634 \pm 0.005$ \\
cardiovascular system disease ({\tt DOID:1287}) & $0.726 \pm 0.005$ \\
disease by infectious agent ({\tt DOID:0050117}) & $0.617 \pm 0.005$ \\
disease of cellular proliferation ({\tt DOID:14566}) & $0.696 \pm 0.005$ \\
disease of mental health ({\tt DOID:150}) & $0.814 \pm 0.004$ \\
disease of metabolism ({\tt DOID:0014667}) & $0.810 \pm 0.004$ \\
endocrine system disease ({\tt DOID:28}) & $0.696 \pm 0.005$ \\
fungal infectious disease ({\tt DOID:1564}) & $0.665 \pm 0.005$ \\
gastrointestinal system disease ({\tt DOID:77}) & $0.704 \pm 0.005$ \\
genetic disease ({\tt DOID:630}) & $0.659 \pm 0.005$ \\
immune system disease ({\tt DOID:2914}) & $0.711 \pm 0.005$ \\
integumentary system disease ({\tt DOID:16}) & $0.696 \pm 0.005$ \\
musculoskeletal system disease ({\tt DOID:17}) & $0.740 \pm 0.005$ \\
nervous system disease ({\tt DOID:863}) & $0.705 \pm 0.005$ \\
parasitic infectious disease ({\tt DOID:1398}) & $0.577 \pm 0.005$ \\
physical disorder ({\tt DOID:0080015}) & $0.611 \pm 0.005$ \\
reproductive system disease ({\tt DOID:15}) & $0.808 \pm 0.004$ \\
respiratory system disease ({\tt DOID:1579}) & $0.855 \pm 0.004$ \\
syndrome ({\tt DOID:225}) & $0.632 \pm 0.005$ \\
thoracic disease ({\tt DOID:0060118}) & $0.927 \pm 0.003$ \\
urinary system disease ({\tt DOID:18}) & $0.817 \pm 0.004$ \\
viral infectious disease ({\tt DOID:934}) & $0.667 \pm 0.005$ \\
\hline
\end{tabular}
\caption{\label{tbl:categories}Phenotypic homogeneity of disease
  categories. We compute ROCAUC values for top-level categories in
  DO. Diseases are ranked based on phenotypic similarity, true
  positive matches are diseases in the same top-level DO category, and
  negative matches are diseases in different DO categories.}
\end{table}

\end{document}